\documentclass[%
 reprint,superscriptaddress,
 amsmath,amssymb,prl,aps,
]{revtex4-2}

\usepackage{graphicx}% Include figure files
\usepackage{dcolumn}% Align table columns on decimal point
\usepackage{bm}% bold math
\usepackage{hyperref}% add hypertext capabilities
\usepackage{xcolor}
\usepackage{mathptmx}

\begin{document}
\preprint{APS/123-QED}

\title{Gaps in Topological Magnon Spectra: Intrinsic vs. Extrinsic Effects} 

\author{Seung-Hwan Do}
\affiliation{Materials Science and Technology Division, Oak Ridge National Laboratory, Oak Ridge, Tennessee 37831, USA}

\author{Joseph A. M. Paddison}
\affiliation{Materials Science and Technology Division, Oak Ridge National Laboratory, Oak Ridge, Tennessee 37831, USA}

\author{Gabriele Sala}
\affiliation{Second Target Station, Oak Ridge National Laboratory, Oak Ridge, Tennessee 37831, USA}

\author{Travis J. Williams}
\affiliation{Neutron Scattering Division, Oak Ridge National Laboratory, Oak Ridge, Tennessee 37831, USA}

\author{Koji Kaneko}
\affiliation{Materials Sciences Research Center, Japan Atomic Energy Agency, Tokai, Ibaraki 319-1195, Japan}
\affiliation{Advanced Science Research Center, Japan Atomic Energy Agency, Tokai, Ibaraki 319-1195, Japan}

\author{Keitaro Kuwahara}
\affiliation{Institute of Quantum Beam Science, Ibaraki University, Mito, Ibaraki 310-8512, Japan}

\author{A. F. May}
\affiliation{Materials Science and Technology Division, Oak Ridge National Laboratory, Oak Ridge, Tennessee 37831, USA}

\author{Jiaqiang Yan}
\affiliation{Materials Science and Technology Division, Oak Ridge National Laboratory, Oak Ridge, Tennessee 37831, USA}

\author{Michael A. McGuire}
\affiliation{Materials Science and Technology Division, Oak Ridge National Laboratory, Oak Ridge, Tennessee 37831, USA}

\author{Matthew B. Stone}
\affiliation{Neutron Scattering Division, Oak Ridge National Laboratory, Oak Ridge, Tennessee 37831, USA}

\author{Mark D. Lumsden}
\affiliation{Neutron Scattering Division, Oak Ridge National Laboratory, Oak Ridge, Tennessee 37831, USA}

\author{Andrew D. Christianson}
\affiliation{Materials Science and Technology Division, Oak Ridge National Laboratory, Oak Ridge, Tennessee 37831, USA}

%\date{\today}% It is always \today, today, but any date may be explicitly specified

\begin{abstract}
For topological magnon spectra, determining and explaining the presence of a gap at a magnon crossing point is a key to characterize the topological properties of the system. An inelastic neutron scattering study of a single crystal is a powerful experimental technique that is widely employed to probe the magnetic excitation spectra of topological materials.
Here, we show that when the scattering intensity rapidly disperses in the vicinity of a crossing point, such as a Dirac point, the apparent topological gap size is extremely sensitive to experimental conditions including sample mosaic, resolution, and momentum integration range. We demonstrate these effects using comprehensive neutron-scattering measurements of CrCl$_3$. Our measurements confirm the gapless nature of the Dirac magnon in CrCl$_3$, but also reveal an artificial, i.e. extrinsic, magnon gap unless the momentum integration range is carefully controlled.  Our study provides an explanation of the apparent discrepancies between spectroscopic and first-principles estimates of Dirac magnon gap sizes, and provides guidelines for accurate spectroscopic measurement of topological magnon gaps.

\end{abstract}

This manuscript has been authored by UT-Battelle, LLC under Contract No. DE-AC05-00OR22725 with the U.S. Department of Energy.  The United States Government retains and the publisher, by accepting the article for publication, acknowledges that the United States Government retains a non-exclusive, paid-up, irrevocable, world-wide license to publish or reproduce the published form of this manuscript, or allow others to do so, for United States Government purposes.  The Department of Energy will provide public access to these results of federally sponsored research in accordance with the DOE Public Access Plan (http://energy.gov/downloads/doe-public-access-plan).

\maketitle

\begin{figure}[t]
\includegraphics[width=8.5cm]{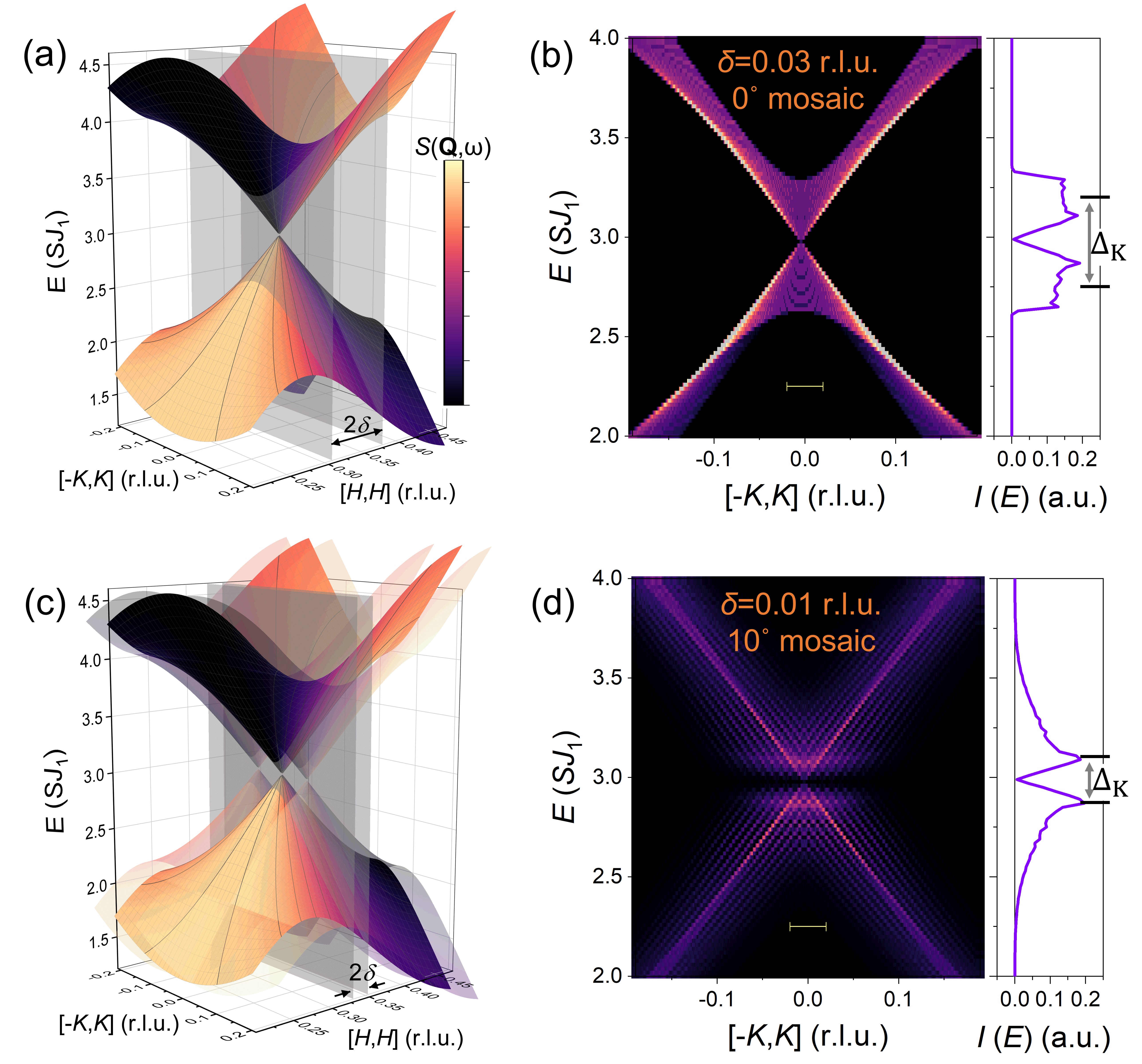} 
\caption{\label{fig:concept} 
Demonstration of how data processing parameters (\textit{Q}-range of integration) and sample mosaicity lead to the appearance of artificial gaps in a gapless Dirac magnon spectrum. 
(a) Gapless Dirac magnon dispersion for spins $S$ on a honeycomb lattice with ferromagnetic nearest-neighbor interaction $J_1$. (b) The \textbf{\textit{Q}}-\textit{E} contour map is obtained by integrating over data within the finite orthogonal \textit{Q}-range, 2$\delta$ shown in (a). Due to rapid dispersion away from the crossing point, an artificial gap ($\Delta_\text{K}$) appears in the spectrum with an apparent size ($\Delta_\text{K}$, defined as the distance between the two peak centers) that depends on the \textbf{\textit{Q}}-integration range.
(c) Sample mosaicity generates a broad distribution of off-centered Dirac cones in \textbf{\textit{Q}}, and the superposed magnon spectra produces an artificial gap at the K-point. (d) Calculated Dirac magnon spectra for a 10$^{\circ}$ sample mosaicity with a $\delta$=0.01 r.l.u.. Data within the yellow bar are integrated to obtain the energy scan at the K-point (far right panels). 
}
\end{figure}

Magnons in a honeycomb-lattice ferromagnet have an analogous description to the single-orbital tight-binding model for electrons in graphene. As a result, the magnons in a honeycomb ferromagnet exhibit a linear energy-momentum dispersion near a magnon crossing point, often termed a Dirac magnon [Fig.~\ref{fig:concept}]~\cite{goerbig2017,owerre2016,pershoguba2018}. Observing the Dirac magnon is of intense current interest as a means of assessing potential topological properties. In particular, if spin-orbit coupling (SOC) accompanying a broken inversion symmetry generates a nonzero Dzyalloshinskii–Moriya interaction (DMI), a gap opens at the Dirac crossing point, in close analogy with the spin-orbit-induced semiconducting gap in graphene. This can yield a topologically-protected magnon edge-state with low-dissipation, which results in a non-trivial topological magnon insulator~\cite{owerre2016,pershoguba2018,chisnell2015,aguilera2020}. Consequently, experimentally identifying materials with a topological magnon gap (TMG) is a crucial goal, with wide-ranging implications for spin-transport-based technologies such as magnon spintronics~\cite{li2021,malz2019,ruckriege2018}.

%%%%%%%%%%%%%%%%%%%%%%Figure 1%%%%%%%%%%%%%%%%%%%%%%%%%%%%%%%%%%%%%%%%%%%%%%%%%%%%%%%%%%%%%%%%%%
%%%%%%%%%%%%%%%%%%%%%%%%%%%%%%%%%%%%%%%%%%%%%%%%%%%%%%%%%%%%%%%%%%%%%%%%%%%%%%%%%%%%%%%%%%%%%%%%

%%%%%%%%%%%%%%%%%%%%%%Figure 2%%%%%%%%%%%%%%%%%%%%%%%%%%%%%%%%%%%%%%%%%%%%%%%%%%%%%%%%%%%%%%%%%%
\begin{figure*}[t]
\includegraphics[width=17cm]{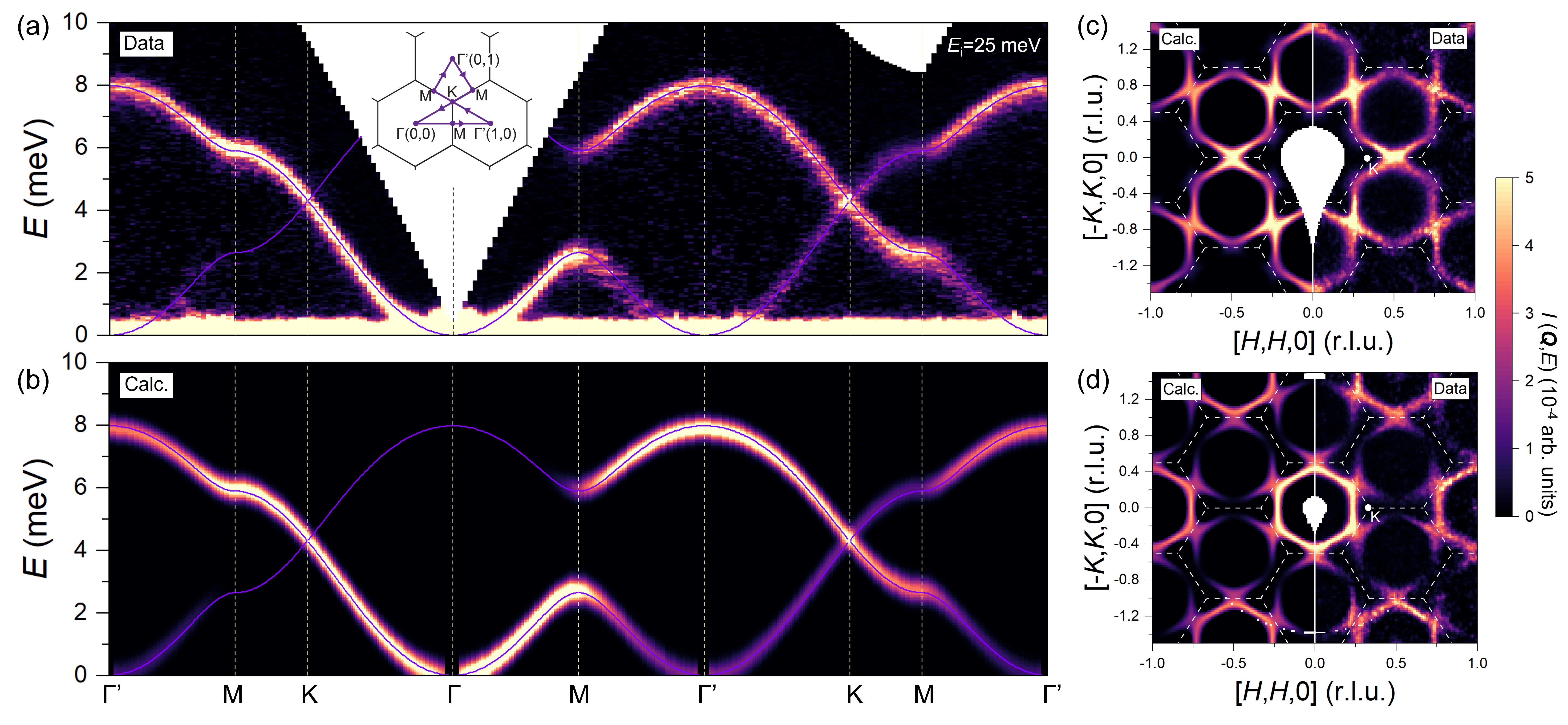} 
\caption{\label{fig:fullspectra} 
(a) INS spectra of CrCl$_{3}$  measured at $T=5$~K ($<T_\text{N}$) along high-symmetry directions as indicated in the $HK$-reciprocal space map shown in the inset. Data were obtained by integrating over a thickness of $Q=$0.024 \AA$^{-1}$  (0.02 r.l.u. along $[H,0,0]$) in the $HK$-plane and $-2.5\leq [0,0,L] \leq2.5$ along the out-of-plane direction. (b) Corresponding spin-wave calculations convolved with the energy resolution function of the instrument. Constant energy slices for (c) $E=6.0\pm0.15$ meV and (d) $E=2.5\pm0.15$ meV are compared with the corresponding calculations.
}
\end{figure*}
%%%%%%%%%%%%%%%%%%%%%%%%%%%%%%%%%%%%%%%%%%%%%%%%%%%%%%%%%%%%%%%%%%%%%%%%%%%%%%%%%%%%%%%%%%%%%%%%

Several chromium-based (Cr$^{3+}$, $S$=3/2) honeycomb ferromagnetic materials have been proposed to host gapped Dirac magnons, namely CrBr$_3$, CrI$_3$, and Cr$M$Te$_3$ ($M$=Si, Ge)~\cite{chen2018,chen2021cri,cai2021,zhu2021}. In all of these materials, the key experimental evidence for a TMG has been obtained from single-crystal neutron spectroscopy experiments, which report gaps $\sim3$ to $5$ meV. However, the origin of the gap opening remains unclear, with proposals including DMI, Kitaev interactions, and magnon-phonon coupling~\cite{chen2018,kvashnin2020,chen2021cri,delugas2021,lee2020,aguilera2020}. Furthermore, the estimated gaps based upon the DMI and Kitaev interaction strengths are much bigger than the predicted gap size of first-principle calculations for the small SOC of half-filled $t_{2g}$ orbitals of Cr$^{3+}$~\cite{kvashnin2020,stavropoulos2021}.
In this context, accurate neutron spectroscopy experiments can be crucial, because they provide a measurement of the magnon dispersion in $4$-dimensional momentum-energy (\textbf{\textit{Q}}-$E$) space that can, in principle, be quantitatively compared with calculations. In practice, however, the weakness of the neutron-sample interaction often requires compromises to increase the signal strength, such as the use of many co-aligned single crystals with increased mosaicity, the relaxation of \textbf{\textit{Q}} and \textit{E} resolution to gain more neutron flux, and the integration of measured data over a significant range of \textbf{\textit{Q}} and \textit{E} to improve counting statistics of the measurement. These approaches to enhance signal can result in spurious findings when the neutron intensity rapidly disperses or has a singularity in \textbf{\textit{Q}}-\textit{E} space as is the case for a Dirac magnon, and this may in turn cause a significant impact on the observation and estimates of the TMG.

In this letter, we present a detailed analysis of extrinsic effects on spectroscopic estimates of TMG sizes. Our key finding is that a dominant \emph{extrinsic} contribution to the apparent magnon gap occurs if ``typical" \textbf{\textit{Q}}-integration ranges are used, which can cause the intrinsic TMG to be substantially overestimated or misdiagnosed. We identify this effect using a simple model. We then present high-resolution neutron spectroscopy measurements on a single crystal of CrCl$_3$. Our results establish CrCl$_3$ as an ideal quasi-2D Heisenberg ferromagnet with a gapless Dirac magnon, and provide a comprehensive map of the excitation spectrum near the Dirac point. However, we also find that an extrinsic magnon gap appears unless the \textbf{\textit{Q}}-integration range is carefully controlled. Finally, we reinterpret published spectroscopic studies of CrBr$_3$ \cite{cai2021} and CrSiTe$_3$ \cite{zhu2021}, and show that the apparent magnon gaps in these materials likely contain substantial extrinsic contributions. Our results resolve apparent discrepancies between spectroscopic and first-principles estimates of TMG sizes, and provide guidelines for accurate spectroscopic measurement of topological magnon gaps.

Figure~\ref{fig:concept} shows calculated magnon spectra for a honeycomb-lattice Heisenberg ferromagnet with nearest-neighbor (n.n.) interaction $J_1$. This model preserves effective time-reversal magnon symmetry and exhibits a gapless Dirac magnon. As shown in Fig.~\ref{fig:concept}(a), two cone-shaped dispersions touch at a single location in three-dimensional \textbf{\textit{Q}}-\textit{E} space. The linear dispersion near the Dirac point is revealed by taking a slice along the $[-K,K,0]$ direction [Fig.~\ref{fig:concept}(b)]. In practice, it is necessary to integrate the intensity over a finite \textbf{\textit{Q}}-range (denoted $2\delta$) orthogonal to the slice direction. The gapless Dirac magnon is faithfully reproduced only in the limit $\delta \rightarrow 0$ [Fig.~\ref{fig:concept}(a)]. In contrast, taking $\delta=0.03$ r.l.u. yields an apparent gap of size $\sim$ $0.5 SJ_1$, which is $8$\% of the full bandwidth of $6 SJ_1$ [Fig.~\ref{fig:concept}(b)]. Here, our choice of $\delta$ is informed by previous experimental studies, in which ``typical" values of $\delta$ are between $0.03$ and $0.20$ r.l.u.~\cite{yuan2020,cai2021,chen2018,chen2021cri}. This extrinsic gap occurs because the intensity is averaged over the Dirac cone away from the Dirac point, and is substantial even for small $\delta$ because the intensity varies rapidly with \textbf{\textit{Q}}. 
This \textbf{\textit{Q}}-integration effect is also distinct from the effect of sample mosaicity such as considered for recent study of CrI$_3$\cite{chen2021cri}, which generates a superposition of off-centered Dirac cones [Fig.~\ref{fig:concept}(c)], and causes an additional artificial gap at the Dirac point [Fig.~\ref{fig:concept}(d)].

We investigate extrinsic contributions to the magnon gap using spectroscopic measurements of the van-der-Waals ferromagnet CrCl$_3$, which contains undistorted Cr$^{3+}$ honeycomb layers below 150 K (space group $R\bar{3}$) ~\cite{mcguire2017,glamazda2017}.
CrCl$_3$ has a ferromagnetic spin alignment in the plane in each honeycomb layer below $T_\text{N}=14$ K~\cite{mcguire2017}. Compared to CrBr$_3$ and CrI$_3$, the relatively light Cl ligand atom of CrCl$_3$ is likely to give a small SOC and hence anisotropic exchange interactions should also be correspondingly weak and host a gapless Dirac magnon, as suggested by recent neutron-scattering measurements on polycrystalline CrCl$_3$ samples~\cite{chen2021crcl,schneeloch2021}. Furthermore, unlike other Cr-halides such as CrI$_3$, CrCl$_3$ can be grown in as a large single crystal with a small sample mosaic. 
Consequently, CrCl$_3$ is an ideal model system to investigate the impact of extrinsic effects and data treatment on the spin-wave spectrum near the Dirac point.

%%%%%%%%%%%%%%%%%%%%%%Figure 3%%%%%%%%%%%%%%%%%%%%%%%%%%%%%%%%%%%%%%%%%%%%%%%%%%%%%%%%%%%%%%%%%%
\begin{figure*}[t]
\includegraphics[width=17cm]{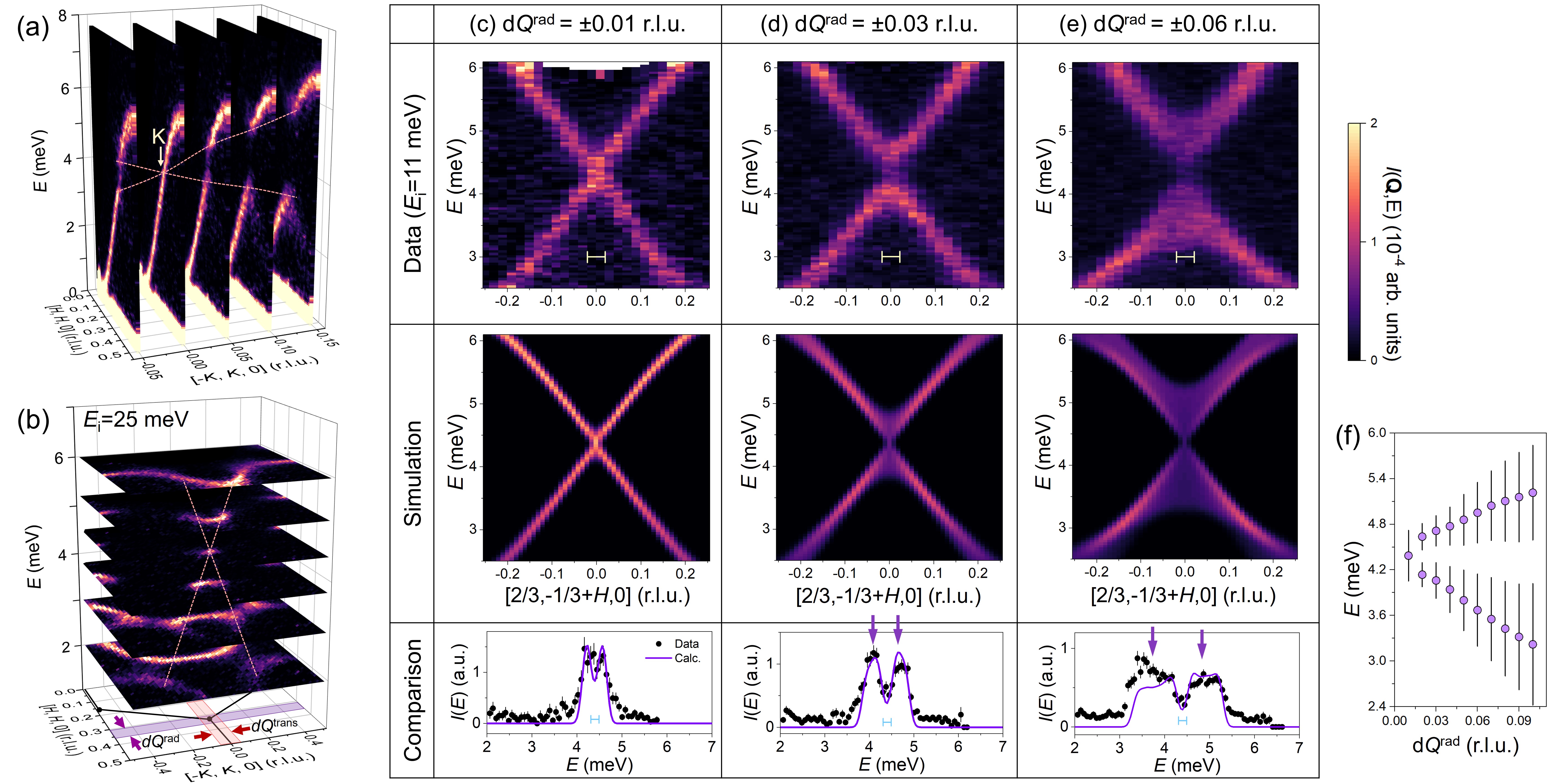}
\caption{\label{fig:Dirac}
(a) $\textbf{\textit{Q}}$-$E$ slices along [$H$, $H$, 0]-directions at \textit{K}$=-0.05$, $0$, $0.05$, $0.1$, and $0.15$ for $[-K,K,0]$. Data were obtained by integrating over $-0.01\leq K \leq 0.01$ and $-2.5\leq L \leq 2.5$. The dashed lines connect ends of acoustic and optical magnon bonds varying with $[-K,K,0]$, which shows a crossing of two magnon bands at a Dirac point, $\mathrm{K} =(\frac{1}{3},\frac{1}{3},0)$. (b) Constant energy slices of the INS spectra. The dashed lines indicate energy evolution of acoustic and optical magnons along the transverse \textbf{\textit{Q}}-direction near the Dirac point. 
(c-d) Opening an artificial gap at crossing point, $(\frac{2}{3},-\frac{1}{3},0)$, with increasing integration range, d${Q}^\text{rad}$.
The spectra were collected using high resolution mode for $E_{i}=11$ meV with the SEQUOIA spectrometer, and obtained by integrating over (c) d${Q}^\text{rad}=\pm0.01$ r.l.u., (d) d${Q}^\text{rad}=\pm0.03$ r.l.u., and (e) d${Q}^\text{rad}=\pm0.06$ r.l.u. ($0.02$, $0.06$, $0.13$\AA$^{-1}$, respectively). The energy-resolution convolved spin-wave spectra are compared in the second and third rows, and are calculated for the identical \textbf{\textit{Q}}-integration range as the experimental data. The constant \textbf{\textit{Q}}-scans at the Dirac point were compared by integrating over d${Q}^\text{trans}=\pm0.02$ r.l.u. (a yellow bar in the contour map). The blue bar is the instrumental resolution at $E^{Dirac}=4.4$ meV. The peak positions and full-width-at-half maximum are quantified by Gaussian fits and exhibited as points with error-bars, respectively, as a function of d${Q}^\text{rad}$ in (f). The distance between the positions indicate the apparent gap size.
}
\end{figure*}
%%%%%%%%%%%%%%%%%%%%%%%%%%%%%%%%%%%%%%%%%%%%%%%%%%%%%%%%%%%%%%%%%%%%%%%%%%%%%%%%%%%%%%%%%%%%%%%%%

A CrCl$_3$ single crystal with a mosaic $<0.68^{\circ}$ was grown by re-crystallization of commercial CrCl$_3$ powder~\cite{may2020}. The INS data was collected with the SEQUOIA time-of-flight spectrometer~\cite{granroth2010} at the Spallation Neutron Source located at Oak Ridge National Laboratory. The Fermi choppers were phased for the \textit{high-resolution} mode of SEQUOIA for incident energies $E_\text{i}=4$, $11$, and $25$ meV~\cite{supple}. Figure~\ref{fig:fullspectra}(a) shows the measured magnon spectra at $5$ K ($T<T_\text{N}$). The spectra are composed of one acoustic and one optical magnon branch with an overall magnon bandwidth of $8$ meV ($W_{HK}$) in the $HK$-plane. The acoustic magnon emanates from the $\Gamma$ point $(0,0,0)$, and meets the optical magnon at the K-points, exhibiting typical spin-waves for a ferromagnetic honeycomb lattice~\cite{owerre2016}.

To model the observed magnon spectra, we used linear spin-wave theory (LSWT) using the SpinW package~\cite{toth2015}. The spin Hamiltonian was modeled with ${\cal H}=J_{n}\sum_{i,j} S_{i}S_{j} -D^{z}\sum_{i}(S_{i}^{z})^{2}$, for $S=3/2$ of Cr$^{3+}$. We consider Heisenberg ($J_{n}$) exchanges up to the third  n.n. interactions in the honeycomb plane and an easy-plane single-ion anisotropy ($D^{z}>0$). 
We exclude bond-dependent exchange interactions (Kitaev and $\Gamma$ terms) because these would yield different mode energies at different K-points coupled with the aligned spin direction~\cite{elliot2020}, in disagreement with our experimental data. We also note that the symmetry-allowed second n.n. DMI term cannot generate a magnon gap at the K-point for the in-plane magnetic structure with the symmetry operations of exchange matrix for $R \bar{3}$ structure~\cite{yuan2020}.
The inter-layer coupling was assumed to be negligible due to the highly two-dimensional spin-wave dispersion~(see \cite{supple} for a sensitivity of interlayer coupling in spin-wave analysis).
The calculated magnon dispersions from the model Hamiltonian were fitted to the experimental dispersion points extracted by fitting a Gaussian function to the INS spectra.  The best fit was for $J_{1}=-0.934(5)$ meV, $J_{2}=-0.0302(2)$ meV, $J_{3}=0.0488(1)$ meV, and $D^{z}=0.0100(1)$ meV~\cite{supple}. The dominant n.n. ferromagnetic interaction $J_{1}$ is responsible for the formation of the ferromagnetic honeycomb spin-lattice in CrCl$_3$. The single-ion anisotropy was found to be nearly zero, indicating a nearly isotropic Cr$^{3+}$ spin. Fig.~\ref{fig:fullspectra}(b-d) show that the calculated spectra reproduce the measured spectra with high fidelity, demonstrating a quasi-two-dimensional Heisenberg ferromagnetic honeycomb spin-lattice for CrCl$_3$. The determined Hamiltonian preserves time-reversal symmetry of magnons, resulting in a gapless Dirac magnon at the K-point.

Details of the magnon spectra near the Dirac point are shown in Fig.~\ref{fig:Dirac}. In the figure, the magnon spectra are displayed along the radial \textbf{\textit{Q}}-direction (parallel to $[H,H,0]$) with varying transverse \textbf{\textit{Q}}-component ($[-K,K,0]$). 
The separate acoustic and optical modes intersect at the single \textbf{\textit{Q}}-position of the Dirac point $(\frac{1}{3},\frac{1}{3},0)$. This sharp band touching is viewed as a nodal point at $E^{Dirac}=4.4$ meV in the horizontal \textbf{\textit{Q}}-$E$ slices of the spectra [Fig.~\ref{fig:Dirac}(b)], representing the shape of a Dirac cone dispersion. Deviating from $E^\text{Dirac}$, the Dirac point evolves into triangular scattering patterns around K [Fig.\ref{fig:fullspectra}(c,d)] with modulating intensity across the Dirac point in energy. 
The variation in intensity is associated with the isospin locked with offset momentum winding around the Dirac point~\cite{elliot2020,shivam2017}. 
As the result, the two conical dispersions having anti-phased winding patterns of intensity meet at the K-point [refer to Fig~\ref{fig:concept}(a)]. When $\textbf{\textit{Q}}$ is perpendicular to the radial-direction, the two bands have an identical magnon structure factor, and the Dirac magnon reveals a clear band crossing, as shown in Fig.~\ref{fig:Dirac}(c), evidencing a gapless Dirac magnon in CrCl$_3$. 

As described above, a focal issue in the search for topological properties of the magnon spectrum is identifying the presence of an intrinsic gap (TMG) at the K-point. Here we investigate the impact of \textit{Q}-integration range on the spectrum near the Dirac point in CrCl$_3$.
Fig.~\ref{fig:Dirac}(c)-(e) show the results of the differently histogrammed spectra in the vicinity of the K-point while varying the orthogonal momentum-integration range, d${Q}^\text{rad}$. (see Fig.~\ref{fig:Dirac}(b) for d${Q}^\text{rad}$).
For d${Q}^\text{rad}=\pm0.01$, the experimental spectrum shows a gapless feature [Fig.\ref{fig:Dirac}(c)]. However, a slightly wider integration range with d${Q}^\text{rad}=\pm0.03$ opens an \textit{apparent} gap with $\Delta_\text{K} \sim 0.7$ meV ($\sim0.09$ $W_{HK}$) [Fig.\ref{fig:Dirac}(d)] and the apparent gap size increases with $\Delta_\text{K} \sim1.3$ meV ($0.16$ $W_{HK}$) for d${Q}^\text{rad}=\pm0.06$ [Fig.\ref{fig:Dirac}(e)] (see Fig.~\ref{fig:Dirac}(f) for d${Q}^\text{rad}$ dependence of the gap size). The measured spectra are directly compared to the spin-wave calculation from the obtained Hamiltonian parameters, reflecting the resolution function of the SEQUOIA spectrometer for the identical momentum integration ranges. The good agreement between calculations and experiment demonstrates that the apparent gap is fully explained by the choice of \textbf{\textit{Q}}-integration range. This artificial gap consistently appears in the radial \textbf{\textit{Q}} scan for d${Q}^\text{trans}$ range, reflecting the conical shape of dispersion. 
It is worth noting that this \textit{apparent} gap exhibits a finite spectral intensity within the gap, where the instrumental resolution is smaller than the apparent gap size, which is in contrast to the TMG case showing zero intensity between the peak splitting~\cite{supple}. Therefore, to clarify TMG, careful comparison of the measured data to the energy-resolution convolved spin-wave calculation including momentum integration range should be performed.

%%%%%%%%%%%%%%%%%%%%%%Figure 4%%%%%%%%%%%%%%%%%%%%%%%%%%%%%%%%%%%%%%%%%%%%%%%%%%%%%%%%%%%%%%%%%%
\begin{figure}[t]
\includegraphics[width=8.5cm]{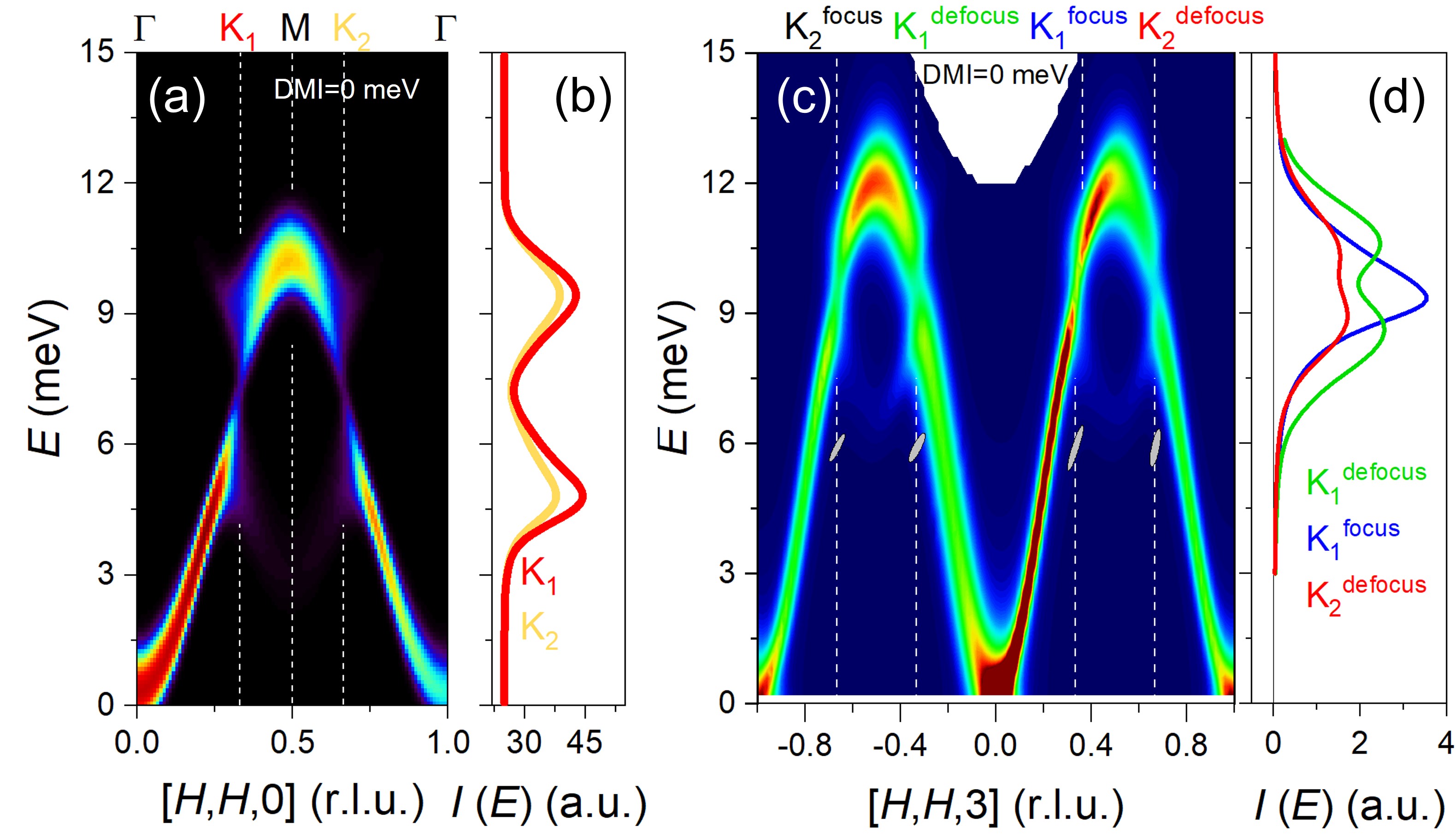} 
\caption{\label{fig:reinterpret} 
Calculations of INS spectra for topological magnon candidates, considering extrinsic effects. (a) Calculated spectra for \textit{time-of-flight} spectrometer experiment for CrBr$_3$~\cite{cai2021}. Spin-waves were calculated for only the Heisenberg Hamiltonian parameters considering the integration range d${Q}^\text{trans}=0.2$ r.l.u. ($0.23$\AA$^{-1}$). (b) Calculated energy scans at K$_1$ and K$_2$. (c) Calculated spectra for a \textit{triple-axis} spectrometer experiment for CrSiTe$_3$~\cite{zhu2021}. The spin-wave spectrum was calculated for only the Heisenberg parameters, and convolved with the instrumental energy resolution function for the experimental conditions presented in Ref.~\cite{zhu2021}. The simulated resolution ellipse functions at the K-points at $E^\text{Dirac}=10$ meV are indicated with gray ellipse. The K-points are divided into `focus' and `defocus' regions according to the coupling status of the dispersion and resolution ellipse, and the energy scans simulated at K$_1^\text{focus}$, K$_1^\text{defocus}$, and K$_2^\text{defocus}$ are compared in (d).
}
\end{figure}
%%%%%%%%%%%%%%%%%%%%%%%%%%%%%%%%%%%%%%%%%%%%%%%%%%%%%%%%%%%%%%%%%%%%%%%%%%%%%%%%%%%%%%%%%%%%%%%%

The discussion above has important implications for the ongoing debate regarding the observations of the Dirac magnon gap in other chromium van der Waals honeycomb ferromagnets. The previously used momentum integration range and a large sample mosaic for CrBr$_3$ (d${Q}^\text{trans}=\pm0.2$) and CrI$_3$ (sample mosaic$=8\sim17^{\circ}$), respectively, are likely to cause a large extrinsic gap contribution~\cite{cai2021,chen2018,chen2021cri}, thus the TMGs are possibly overestimated.
Figure~\ref{fig:reinterpret}(a) shows the calculated INS spectrum of the spin-wave spectra for CrBr$_3$~\cite{cai2021}, including the \textit{Q}-integration range and instrumental resolution used in Ref.~\cite{cai2021}. This calculation assumes \textit{only} Heisenberg interactions ($J_1=-1.48$ meV, $J_2=-0.08$ meV, $J_3=0.11$, and $D^{z}=-0.02$ meV, from Ref.~\cite{cai2021}) with zero DMI. For simplicity, the sample mosaic was assumed to be zero. 
The resulting spectrum shows the upper and lower magnon spectra having decreased intensities near the Dirac points (K$_1$, K$_2$), which reproduces the observed magnon dispersion in Ref.~\cite{cai2021}. Noticeably, most of the observed gap, corresponding to the peak splittings $\sim4$ meV, in the energy scans at K$_1$ and K$_2$ is explained by the orthogonal momentum integration range effect on a pure Heisenberg model, without DMI. As a consequence, the size of the gap at the Dirac point is likely overestimated.

In contrast to time-of-flight spectroscopy measurement on Cr trihalides, topological magnon candidtates Cr$M$Te$_3$ ($M$=Si,Ge) were measured using triple-axis spectrometers~\cite{zhu2021}. For triple-axis measurements, integration ranges are not a concern but resolution effects can be significant, particularly when horizontal focusing is used. Therefore, the observed spectra are strongly coupled with the shape of the resolution function (gray ellipse in Fig.~\ref{fig:reinterpret}(c)). Figure~\ref{fig:reinterpret}(c) exhibits newly calculated spin-wave spectra for CrSiTe$_3$~\cite{zhu2021}, including the resolution calculation for the experiment using Reslib~\cite{reslib}. For the spin-wave calculation, we used \textit{only} Heisenberg spin Hamiltonian parameters ($J_1=-1.49$ meV, $J_2=-0.15$ meV, $J_{c1}=-0.07$, $J_{c2}=-0.06$, and $D^{z}=-0.01$ meV, from Ref.~\cite{zhu2021}) with zero DMI. 
As can be seen, while the gapless Dirac dispersion parallel to the the ellipse (focused) at K$_1^\text{focus}$ leads to a single peak (gapless), the Dirac dispersions anti-parallel to the ellipse (defocused) at K$_1^\text{defocus}$ and K$_2^\text{defocus}$ lead to an apparent gaps in the constant \textbf{\textit{Q}}-scans [Fig.~\ref{fig:reinterpret}(d)]. In this case, the defocused resolution condition plays similar role to the orthogonal \textbf{\textit{Q}}-integration range in Fig.~\ref{fig:Dirac}. 
Confirming the TMG in Cr$M$Te$_3$ ($M$=Si,Ge) requires quantitative comparison of the data and the fully resolution-convolved simulation. Assuming that the data in Ref.~\cite{zhu2021} (Fig.2H) was measured at the focused K$_1$ $(\frac{1}{3},\frac{1}{3},3)$ and defocused K$_2$ $(\frac{2}{3},\frac{2}{3},3)$, our new resolution-convolved spin-wave simulations suggest a DMI$\sim0.06$ meV that is a half of the value reported in Ref.~\cite{zhu2021} (see \cite{supple} for details).

In conclusion, revealing and understanding topological magnons are important steps towards the realization of magnon-based electronic devices as well as for the fundamental goal of discovering a topological magnon insulator.  To meet these challenges, accurately clarifying and estimating an intrinsic Dirac magnon gap is a critical issue in defining the topological properties, and INS experiments play a key role in this endeavor. In particular, our study provides important guidance for spectroscopic measurements of systems having a singularity or a band crossing where the spectrum rapidly disperses in momentum space. 
We have shown that instrumental and data-processing effects can introduce an artificial gap, so that accurate estimation of the topological gap size requires careful data histograms and comparison with resolution-convolved calculations. 
Our results are relevant not only to Dirac magnon gaps, but also to similar Weyl magnon, determining avoided crossing of rattler modes to acoustic phonons~\cite{christensen2008}, and quantifying the life-time of rapidly decaying phonon spectra~\cite{xiao2022}, where similar effects may be anticipated.

\begin{acknowledgments}
We thank Yixi Su for providing their experimental conditions and  Michael E. Manley for useful discussions. This research was supported by the U.S. Department of Energy, Office of Science, Basic Energy Sciences, Materials Science and Engineering Division. Work at the Oak Ridge National Laboratory Spallation Neutron Source was supported by U.S. DOE, Office of Science, BES, Scientific User Facilities Division.
\end{acknowledgments}

\begin{appendix}

\end{appendix}

%\nocite{*}
%\bibliography{mainscript}% Produces the bibliography via BibTeX.

\begin{thebibliography}{30}%
\makeatletter
\providecommand \@ifxundefined [1]{%
 \@ifx{#1\undefined}
}%
\providecommand \@ifnum [1]{%
 \ifnum #1\expandafter \@firstoftwo
 \else \expandafter \@secondoftwo
 \fi
}%
\providecommand \@ifx [1]{%
 \ifx #1\expandafter \@firstoftwo
 \else \expandafter \@secondoftwo
 \fi
}%
\providecommand \natexlab [1]{#1}%
\providecommand \enquote  [1]{``#1''}%
\providecommand \bibnamefont  [1]{#1}%
\providecommand \bibfnamefont [1]{#1}%
\providecommand \citenamefont [1]{#1}%
\providecommand \href@noop [0]{\@secondoftwo}%
\providecommand \href [0]{\begingroup \@sanitize@url \@href}%
\providecommand \@href[1]{\@@startlink{#1}\@@href}%
\providecommand \@@href[1]{\endgroup#1\@@endlink}%
\providecommand \@sanitize@url [0]{\catcode `\\12\catcode `\$12\catcode
  `\&12\catcode `\#12\catcode `\^12\catcode `\_12\catcode `\%12\relax}%
\providecommand \@@startlink[1]{}%
\providecommand \@@endlink[0]{}%
\providecommand \url  [0]{\begingroup\@sanitize@url \@url }%
\providecommand \@url [1]{\endgroup\@href {#1}{\urlprefix }}%
\providecommand \urlprefix  [0]{URL }%
\providecommand \Eprint [0]{\href }%
\providecommand \doibase [0]{http://dx.doi.org/}%
\providecommand \selectlanguage [0]{\@gobble}%
\providecommand \bibinfo  [0]{\@secondoftwo}%
\providecommand \bibfield  [0]{\@secondoftwo}%
\providecommand \translation [1]{[#1]}%
\providecommand \BibitemOpen [0]{}%
\providecommand \bibitemStop [0]{}%
\providecommand \bibitemNoStop [0]{.\EOS\space}%
\providecommand \EOS [0]{\spacefactor3000\relax}%
\providecommand \BibitemShut  [1]{\csname bibitem#1\endcsname}%
\let\auto@bib@innerbib\@empty
%</preamble>
\bibitem [{\citenamefont {Goerbig}\ and\ \citenamefont
  {Montambaux}(2017)}]{goerbig2017}%
  \BibitemOpen
  \bibfield  {author} {\bibinfo {author} {\bibfnamefont {M.}~\bibnamefont
  {Goerbig}}\ and\ \bibinfo {author} {\bibfnamefont {G.}~\bibnamefont
  {Montambaux}},\ }\enquote {\bibinfo {title} {Dirac fermions in condensed
  matter and beyond},}\ in\ \href {\doibase 10.1007/978-3-319-32536-1_2} {\emph
  {\bibinfo {booktitle} {Dirac Matter}}},\ \bibinfo {editor} {edited by\
  \bibinfo {editor} {\bibfnamefont {B.}~\bibnamefont {Duplantier}}, \bibinfo
  {editor} {\bibfnamefont {V.}~\bibnamefont {Rivasseau}}, \ and\ \bibinfo
  {editor} {\bibfnamefont {J.-N.}\ \bibnamefont {Fuchs}}}\ (\bibinfo
  {publisher} {Springer International Publishing},\ \bibinfo {address} {Cham},\
  \bibinfo {year} {2017})\ pp.\ \bibinfo {pages} {25--53}\BibitemShut {NoStop}%
\bibitem [{\citenamefont {Owerre}(2016)}]{owerre2016}%
  \BibitemOpen
  \bibfield  {author} {\bibinfo {author} {\bibfnamefont {S.~A.}\ \bibnamefont
  {Owerre}},\ }\href {\doibase 10.1088/0953-8984/28/38/386001} {\bibfield
  {journal} {\bibinfo  {journal} {Journal of Physics: Condensed Matter}\
  }\textbf {\bibinfo {volume} {28}},\ \bibinfo {pages} {386001} (\bibinfo
  {year} {2016})}\BibitemShut {NoStop}%
\bibitem [{\citenamefont {Pershoguba}\ \emph {et~al.}(2018)\citenamefont
  {Pershoguba}, \citenamefont {Banerjee}, \citenamefont {Lashley},
  \citenamefont {Park}, \citenamefont {\AA{}gren}, \citenamefont {Aeppli},\
  and\ \citenamefont {Balatsky}}]{pershoguba2018}%
  \BibitemOpen
  \bibfield  {author} {\bibinfo {author} {\bibfnamefont {S.~S.}\ \bibnamefont
  {Pershoguba}}, \bibinfo {author} {\bibfnamefont {S.}~\bibnamefont
  {Banerjee}}, \bibinfo {author} {\bibfnamefont {J.~C.}\ \bibnamefont
  {Lashley}}, \bibinfo {author} {\bibfnamefont {J.}~\bibnamefont {Park}},
  \bibinfo {author} {\bibfnamefont {H.}~\bibnamefont {\AA{}gren}}, \bibinfo
  {author} {\bibfnamefont {G.}~\bibnamefont {Aeppli}}, \ and\ \bibinfo {author}
  {\bibfnamefont {A.~V.}\ \bibnamefont {Balatsky}},\ }\href {\doibase
  10.1103/PhysRevX.8.011010} {\bibfield  {journal} {\bibinfo  {journal} {Phys.
  Rev. X}\ }\textbf {\bibinfo {volume} {8}},\ \bibinfo {pages} {011010}
  (\bibinfo {year} {2018})}\BibitemShut {NoStop}%
\bibitem [{\citenamefont {Chisnell}\ \emph {et~al.}(2015)\citenamefont
  {Chisnell}, \citenamefont {Helton}, \citenamefont {Freedman}, \citenamefont
  {Singh}, \citenamefont {Bewley}, \citenamefont {Nocera},\ and\ \citenamefont
  {Lee}}]{chisnell2015}%
  \BibitemOpen
  \bibfield  {author} {\bibinfo {author} {\bibfnamefont {R.}~\bibnamefont
  {Chisnell}}, \bibinfo {author} {\bibfnamefont {J.~S.}\ \bibnamefont
  {Helton}}, \bibinfo {author} {\bibfnamefont {D.~E.}\ \bibnamefont
  {Freedman}}, \bibinfo {author} {\bibfnamefont {D.~K.}\ \bibnamefont {Singh}},
  \bibinfo {author} {\bibfnamefont {R.~I.}\ \bibnamefont {Bewley}}, \bibinfo
  {author} {\bibfnamefont {D.~G.}\ \bibnamefont {Nocera}}, \ and\ \bibinfo
  {author} {\bibfnamefont {Y.~S.}\ \bibnamefont {Lee}},\ }\href {\doibase
  10.1103/PhysRevLett.115.147201} {\bibfield  {journal} {\bibinfo  {journal}
  {Phys. Rev. Lett.}\ }\textbf {\bibinfo {volume} {115}},\ \bibinfo {pages}
  {147201} (\bibinfo {year} {2015})}\BibitemShut {NoStop}%
\bibitem [{\citenamefont {Aguilera}\ \emph {et~al.}(2020)\citenamefont
  {Aguilera}, \citenamefont {Jaeschke-Ubiergo}, \citenamefont {Vidal-Silva},
  \citenamefont {Torres},\ and\ \citenamefont {Nunez}}]{aguilera2020}%
  \BibitemOpen
  \bibfield  {author} {\bibinfo {author} {\bibfnamefont {E.}~\bibnamefont
  {Aguilera}}, \bibinfo {author} {\bibfnamefont {R.}~\bibnamefont
  {Jaeschke-Ubiergo}}, \bibinfo {author} {\bibfnamefont {N.}~\bibnamefont
  {Vidal-Silva}}, \bibinfo {author} {\bibfnamefont {L.~E. F.~F.}\ \bibnamefont
  {Torres}}, \ and\ \bibinfo {author} {\bibfnamefont {A.~S.}\ \bibnamefont
  {Nunez}},\ }\href {\doibase 10.1103/PhysRevB.102.024409} {\bibfield
  {journal} {\bibinfo  {journal} {Phys. Rev. B}\ }\textbf {\bibinfo {volume}
  {102}},\ \bibinfo {pages} {024409} (\bibinfo {year} {2020})}\BibitemShut
  {NoStop}%
\bibitem [{\citenamefont {Li}\ \emph {et~al.}(2021)\citenamefont {Li},
  \citenamefont {Cao},\ and\ \citenamefont {Yan}}]{li2021}%
  \BibitemOpen
  \bibfield  {author} {\bibinfo {author} {\bibfnamefont {Z.-X.}\ \bibnamefont
  {Li}}, \bibinfo {author} {\bibfnamefont {Y.}~\bibnamefont {Cao}}, \ and\
  \bibinfo {author} {\bibfnamefont {P.}~\bibnamefont {Yan}},\ }\href {\doibase
  https://doi.org/10.1016/j.physrep.2021.02.003} {\bibfield  {journal}
  {\bibinfo  {journal} {Physics Reports}\ }\textbf {\bibinfo {volume} {915}},\
  \bibinfo {pages} {1} (\bibinfo {year} {2021})},\ \bibinfo {note} {topological
  insulators and semimetals in classical magnetic systems}\BibitemShut
  {NoStop}%
\bibitem [{\citenamefont {Malz}\ \emph {et~al.}(2019)\citenamefont {Malz},
  \citenamefont {Knolle},\ and\ \citenamefont {Nunnenkamp}}]{malz2019}%
  \BibitemOpen
  \bibfield  {author} {\bibinfo {author} {\bibfnamefont {D.}~\bibnamefont
  {Malz}}, \bibinfo {author} {\bibfnamefont {J.}~\bibnamefont {Knolle}}, \ and\
  \bibinfo {author} {\bibfnamefont {A.}~\bibnamefont {Nunnenkamp}},\
  }\href@noop {} {\bibfield  {journal} {\bibinfo  {journal} {Nature
  communications}\ }\textbf {\bibinfo {volume} {10}},\ \bibinfo {pages} {1}
  (\bibinfo {year} {2019})}\BibitemShut {NoStop}%
\bibitem [{\citenamefont {R\"uckriegel}\ \emph {et~al.}(2018)\citenamefont
  {R\"uckriegel}, \citenamefont {Brataas},\ and\ \citenamefont
  {Duine}}]{ruckriege2018}%
  \BibitemOpen
  \bibfield  {author} {\bibinfo {author} {\bibfnamefont {A.}~\bibnamefont
  {R\"uckriegel}}, \bibinfo {author} {\bibfnamefont {A.}~\bibnamefont
  {Brataas}}, \ and\ \bibinfo {author} {\bibfnamefont {R.~A.}\ \bibnamefont
  {Duine}},\ }\href {\doibase 10.1103/PhysRevB.97.081106} {\bibfield  {journal}
  {\bibinfo  {journal} {Phys. Rev. B}\ }\textbf {\bibinfo {volume} {97}},\
  \bibinfo {pages} {081106} (\bibinfo {year} {2018})}\BibitemShut {NoStop}%
\bibitem [{\citenamefont {Chen}\ \emph {et~al.}(2018)\citenamefont {Chen},
  \citenamefont {Chung}, \citenamefont {Gao}, \citenamefont {Chen},
  \citenamefont {Stone}, \citenamefont {Kolesnikov}, \citenamefont {Huang},\
  and\ \citenamefont {Dai}}]{chen2018}%
  \BibitemOpen
  \bibfield  {author} {\bibinfo {author} {\bibfnamefont {L.}~\bibnamefont
  {Chen}}, \bibinfo {author} {\bibfnamefont {J.-H.}\ \bibnamefont {Chung}},
  \bibinfo {author} {\bibfnamefont {B.}~\bibnamefont {Gao}}, \bibinfo {author}
  {\bibfnamefont {T.}~\bibnamefont {Chen}}, \bibinfo {author} {\bibfnamefont
  {M.~B.}\ \bibnamefont {Stone}}, \bibinfo {author} {\bibfnamefont {A.~I.}\
  \bibnamefont {Kolesnikov}}, \bibinfo {author} {\bibfnamefont
  {Q.}~\bibnamefont {Huang}}, \ and\ \bibinfo {author} {\bibfnamefont
  {P.}~\bibnamefont {Dai}},\ }\href {\doibase 10.1103/PhysRevX.8.041028}
  {\bibfield  {journal} {\bibinfo  {journal} {Phys. Rev. X}\ }\textbf {\bibinfo
  {volume} {8}},\ \bibinfo {pages} {041028} (\bibinfo {year}
  {2018})}\BibitemShut {NoStop}%
\bibitem [{\citenamefont {Chen}\ \emph
  {et~al.}(2021{\natexlab{a}})\citenamefont {Chen}, \citenamefont {Chung},
  \citenamefont {Stone}, \citenamefont {Kolesnikov}, \citenamefont {Winn},
  \citenamefont {Garlea}, \citenamefont {Abernathy}, \citenamefont {Gao},
  \citenamefont {Augustin}, \citenamefont {Santos},\ and\ \citenamefont
  {Dai}}]{chen2021cri}%
  \BibitemOpen
  \bibfield  {author} {\bibinfo {author} {\bibfnamefont {L.}~\bibnamefont
  {Chen}}, \bibinfo {author} {\bibfnamefont {J.-H.}\ \bibnamefont {Chung}},
  \bibinfo {author} {\bibfnamefont {M.~B.}\ \bibnamefont {Stone}}, \bibinfo
  {author} {\bibfnamefont {A.~I.}\ \bibnamefont {Kolesnikov}}, \bibinfo
  {author} {\bibfnamefont {B.}~\bibnamefont {Winn}}, \bibinfo {author}
  {\bibfnamefont {V.~O.}\ \bibnamefont {Garlea}}, \bibinfo {author}
  {\bibfnamefont {D.~L.}\ \bibnamefont {Abernathy}}, \bibinfo {author}
  {\bibfnamefont {B.}~\bibnamefont {Gao}}, \bibinfo {author} {\bibfnamefont
  {M.}~\bibnamefont {Augustin}}, \bibinfo {author} {\bibfnamefont {E.~J.~G.}\
  \bibnamefont {Santos}}, \ and\ \bibinfo {author} {\bibfnamefont
  {P.}~\bibnamefont {Dai}},\ }\href {\doibase 10.1103/PhysRevX.11.031047}
  {\bibfield  {journal} {\bibinfo  {journal} {Phys. Rev. X}\ }\textbf {\bibinfo
  {volume} {11}},\ \bibinfo {pages} {031047} (\bibinfo {year}
  {2021}{\natexlab{a}})}\BibitemShut {NoStop}%
\bibitem [{\citenamefont {Cai}\ \emph {et~al.}(2021)\citenamefont {Cai},
  \citenamefont {Bao}, \citenamefont {Gu}, \citenamefont {Gao}, \citenamefont
  {Ma}, \citenamefont {Shangguan}, \citenamefont {Si}, \citenamefont {Dong},
  \citenamefont {Wang}, \citenamefont {Wu}, \citenamefont {Lin}, \citenamefont
  {Wang}, \citenamefont {Ran}, \citenamefont {Li}, \citenamefont {Adroja},
  \citenamefont {Xi}, \citenamefont {Yu}, \citenamefont {Wu}, \citenamefont
  {Li},\ and\ \citenamefont {Wen}}]{cai2021}%
  \BibitemOpen
  \bibfield  {author} {\bibinfo {author} {\bibfnamefont {Z.}~\bibnamefont
  {Cai}}, \bibinfo {author} {\bibfnamefont {S.}~\bibnamefont {Bao}}, \bibinfo
  {author} {\bibfnamefont {Z.-L.}\ \bibnamefont {Gu}}, \bibinfo {author}
  {\bibfnamefont {Y.-P.}\ \bibnamefont {Gao}}, \bibinfo {author} {\bibfnamefont
  {Z.}~\bibnamefont {Ma}}, \bibinfo {author} {\bibfnamefont {Y.}~\bibnamefont
  {Shangguan}}, \bibinfo {author} {\bibfnamefont {W.}~\bibnamefont {Si}},
  \bibinfo {author} {\bibfnamefont {Z.-Y.}\ \bibnamefont {Dong}}, \bibinfo
  {author} {\bibfnamefont {W.}~\bibnamefont {Wang}}, \bibinfo {author}
  {\bibfnamefont {Y.}~\bibnamefont {Wu}}, \bibinfo {author} {\bibfnamefont
  {D.}~\bibnamefont {Lin}}, \bibinfo {author} {\bibfnamefont {J.}~\bibnamefont
  {Wang}}, \bibinfo {author} {\bibfnamefont {K.}~\bibnamefont {Ran}}, \bibinfo
  {author} {\bibfnamefont {S.}~\bibnamefont {Li}}, \bibinfo {author}
  {\bibfnamefont {D.}~\bibnamefont {Adroja}}, \bibinfo {author} {\bibfnamefont
  {X.}~\bibnamefont {Xi}}, \bibinfo {author} {\bibfnamefont {S.-L.}\
  \bibnamefont {Yu}}, \bibinfo {author} {\bibfnamefont {X.}~\bibnamefont {Wu}},
  \bibinfo {author} {\bibfnamefont {J.-X.}\ \bibnamefont {Li}}, \ and\ \bibinfo
  {author} {\bibfnamefont {J.}~\bibnamefont {Wen}},\ }\href {\doibase
  10.1103/PhysRevB.104.L020402} {\bibfield  {journal} {\bibinfo  {journal}
  {Phys. Rev. B}\ }\textbf {\bibinfo {volume} {104}},\ \bibinfo {pages}
  {L020402} (\bibinfo {year} {2021})}\BibitemShut {NoStop}%
\bibitem [{\citenamefont {Zhu}\ \emph {et~al.}(2021)\citenamefont {Zhu},
  \citenamefont {Zhang}, \citenamefont {Wang}, \citenamefont {dos Santos},
  \citenamefont {Song}, \citenamefont {Mueller}, \citenamefont {Schmalzl},
  \citenamefont {Schmidt}, \citenamefont {Ivanov}, \citenamefont {Park},
  \citenamefont {Xu}, \citenamefont {Ma}, \citenamefont {Lounis}, \citenamefont
  {Blügel}, \citenamefont {Mokrousov}, \citenamefont {Su},\ and\ \citenamefont
  {Brückel}}]{zhu2021}%
  \BibitemOpen
  \bibfield  {author} {\bibinfo {author} {\bibfnamefont {F.}~\bibnamefont
  {Zhu}}, \bibinfo {author} {\bibfnamefont {L.}~\bibnamefont {Zhang}}, \bibinfo
  {author} {\bibfnamefont {X.}~\bibnamefont {Wang}}, \bibinfo {author}
  {\bibfnamefont {F.~J.}\ \bibnamefont {dos Santos}}, \bibinfo {author}
  {\bibfnamefont {J.}~\bibnamefont {Song}}, \bibinfo {author} {\bibfnamefont
  {T.}~\bibnamefont {Mueller}}, \bibinfo {author} {\bibfnamefont
  {K.}~\bibnamefont {Schmalzl}}, \bibinfo {author} {\bibfnamefont {W.~F.}\
  \bibnamefont {Schmidt}}, \bibinfo {author} {\bibfnamefont {A.}~\bibnamefont
  {Ivanov}}, \bibinfo {author} {\bibfnamefont {J.~T.}\ \bibnamefont {Park}},
  \bibinfo {author} {\bibfnamefont {J.}~\bibnamefont {Xu}}, \bibinfo {author}
  {\bibfnamefont {J.}~\bibnamefont {Ma}}, \bibinfo {author} {\bibfnamefont
  {S.}~\bibnamefont {Lounis}}, \bibinfo {author} {\bibfnamefont
  {S.}~\bibnamefont {Blügel}}, \bibinfo {author} {\bibfnamefont
  {Y.}~\bibnamefont {Mokrousov}}, \bibinfo {author} {\bibfnamefont
  {Y.}~\bibnamefont {Su}}, \ and\ \bibinfo {author} {\bibfnamefont
  {T.}~\bibnamefont {Brückel}},\ }\href {\doibase 10.1126/sciadv.abi7532}
  {\bibfield  {journal} {\bibinfo  {journal} {Science Advances}\ }\textbf
  {\bibinfo {volume} {7}},\ \bibinfo {pages} {eabi7532} (\bibinfo {year}
  {2021})}\BibitemShut {NoStop}%
\bibitem [{\citenamefont {Kvashnin}\ \emph {et~al.}(2020)\citenamefont
  {Kvashnin}, \citenamefont {Bergman}, \citenamefont {Lichtenstein},\ and\
  \citenamefont {Katsnelson}}]{kvashnin2020}%
  \BibitemOpen
  \bibfield  {author} {\bibinfo {author} {\bibfnamefont {Y.~O.}\ \bibnamefont
  {Kvashnin}}, \bibinfo {author} {\bibfnamefont {A.}~\bibnamefont {Bergman}},
  \bibinfo {author} {\bibfnamefont {A.~I.}\ \bibnamefont {Lichtenstein}}, \
  and\ \bibinfo {author} {\bibfnamefont {M.~I.}\ \bibnamefont {Katsnelson}},\
  }\href {\doibase 10.1103/PhysRevB.102.115162} {\bibfield  {journal} {\bibinfo
   {journal} {Phys. Rev. B}\ }\textbf {\bibinfo {volume} {102}},\ \bibinfo
  {pages} {115162} (\bibinfo {year} {2020})}\BibitemShut {NoStop}%
\bibitem [{\citenamefont {Delugas}\ \emph {et~al.}(2021)\citenamefont
  {Delugas}, \citenamefont {Baseggio}, \citenamefont {Timrov}, \citenamefont
  {Baroni},\ and\ \citenamefont {Gorni}}]{delugas2021}%
  \BibitemOpen
  \bibfield  {author} {\bibinfo {author} {\bibfnamefont {P.}~\bibnamefont
  {Delugas}}, \bibinfo {author} {\bibfnamefont {O.}~\bibnamefont {Baseggio}},
  \bibinfo {author} {\bibfnamefont {I.}~\bibnamefont {Timrov}}, \bibinfo
  {author} {\bibfnamefont {S.}~\bibnamefont {Baroni}}, \ and\ \bibinfo {author}
  {\bibfnamefont {T.}~\bibnamefont {Gorni}},\ }\href
  {https://arxiv.org/abs/2105.04531} {\bibfield  {journal} {\bibinfo  {journal}
  {arXiv preprint arXiv:2105.04531}\ } (\bibinfo {year} {2021})}\BibitemShut
  {NoStop}%
\bibitem [{\citenamefont {Lee}\ \emph {et~al.}(2020)\citenamefont {Lee},
  \citenamefont {Utermohlen}, \citenamefont {Weber}, \citenamefont {Hwang},
  \citenamefont {Zhang}, \citenamefont {van Tol}, \citenamefont {Goldberger},
  \citenamefont {Trivedi},\ and\ \citenamefont {Hammel}}]{lee2020}%
  \BibitemOpen
  \bibfield  {author} {\bibinfo {author} {\bibfnamefont {I.}~\bibnamefont
  {Lee}}, \bibinfo {author} {\bibfnamefont {F.~G.}\ \bibnamefont {Utermohlen}},
  \bibinfo {author} {\bibfnamefont {D.}~\bibnamefont {Weber}}, \bibinfo
  {author} {\bibfnamefont {K.}~\bibnamefont {Hwang}}, \bibinfo {author}
  {\bibfnamefont {C.}~\bibnamefont {Zhang}}, \bibinfo {author} {\bibfnamefont
  {J.}~\bibnamefont {van Tol}}, \bibinfo {author} {\bibfnamefont {J.~E.}\
  \bibnamefont {Goldberger}}, \bibinfo {author} {\bibfnamefont
  {N.}~\bibnamefont {Trivedi}}, \ and\ \bibinfo {author} {\bibfnamefont
  {P.~C.}\ \bibnamefont {Hammel}},\ }\href {\doibase
  10.1103/PhysRevLett.124.017201} {\bibfield  {journal} {\bibinfo  {journal}
  {Phys. Rev. Lett.}\ }\textbf {\bibinfo {volume} {124}},\ \bibinfo {pages}
  {017201} (\bibinfo {year} {2020})}\BibitemShut {NoStop}%
\bibitem [{\citenamefont {Stavropoulos}\ \emph {et~al.}(2021)\citenamefont
  {Stavropoulos}, \citenamefont {Liu},\ and\ \citenamefont
  {Kee}}]{stavropoulos2021}%
  \BibitemOpen
  \bibfield  {author} {\bibinfo {author} {\bibfnamefont {P.~P.}\ \bibnamefont
  {Stavropoulos}}, \bibinfo {author} {\bibfnamefont {X.}~\bibnamefont {Liu}}, \
  and\ \bibinfo {author} {\bibfnamefont {H.-Y.}\ \bibnamefont {Kee}},\ }\href
  {\doibase 10.1103/PhysRevResearch.3.013216} {\bibfield  {journal} {\bibinfo
  {journal} {Phys. Rev. Research}\ }\textbf {\bibinfo {volume} {3}},\ \bibinfo
  {pages} {013216} (\bibinfo {year} {2021})}\BibitemShut {NoStop}%
\bibitem [{\citenamefont {Yuan}\ \emph {et~al.}(2020)\citenamefont {Yuan},
  \citenamefont {Khait}, \citenamefont {Shu}, \citenamefont {Chou},
  \citenamefont {Stone}, \citenamefont {Clancy}, \citenamefont {Paramekanti},\
  and\ \citenamefont {Kim}}]{yuan2020}%
  \BibitemOpen
  \bibfield  {author} {\bibinfo {author} {\bibfnamefont {B.}~\bibnamefont
  {Yuan}}, \bibinfo {author} {\bibfnamefont {I.}~\bibnamefont {Khait}},
  \bibinfo {author} {\bibfnamefont {G.-J.}\ \bibnamefont {Shu}}, \bibinfo
  {author} {\bibfnamefont {F.~C.}\ \bibnamefont {Chou}}, \bibinfo {author}
  {\bibfnamefont {M.~B.}\ \bibnamefont {Stone}}, \bibinfo {author}
  {\bibfnamefont {J.~P.}\ \bibnamefont {Clancy}}, \bibinfo {author}
  {\bibfnamefont {A.}~\bibnamefont {Paramekanti}}, \ and\ \bibinfo {author}
  {\bibfnamefont {Y.-J.}\ \bibnamefont {Kim}},\ }\href {\doibase
  10.1103/PhysRevX.10.011062} {\bibfield  {journal} {\bibinfo  {journal} {Phys.
  Rev. X}\ }\textbf {\bibinfo {volume} {10}},\ \bibinfo {pages} {011062}
  (\bibinfo {year} {2020})}\BibitemShut {NoStop}%
\bibitem [{\citenamefont {McGuire}\ \emph {et~al.}(2017)\citenamefont
  {McGuire}, \citenamefont {Clark}, \citenamefont {KC}, \citenamefont {Chance},
  \citenamefont {Jellison}, \citenamefont {Cooper}, \citenamefont {Xu},\ and\
  \citenamefont {Sales}}]{mcguire2017}%
  \BibitemOpen
  \bibfield  {author} {\bibinfo {author} {\bibfnamefont {M.~A.}\ \bibnamefont
  {McGuire}}, \bibinfo {author} {\bibfnamefont {G.}~\bibnamefont {Clark}},
  \bibinfo {author} {\bibfnamefont {S.}~\bibnamefont {KC}}, \bibinfo {author}
  {\bibfnamefont {W.~M.}\ \bibnamefont {Chance}}, \bibinfo {author}
  {\bibfnamefont {G.~E.}\ \bibnamefont {Jellison}}, \bibinfo {author}
  {\bibfnamefont {V.~R.}\ \bibnamefont {Cooper}}, \bibinfo {author}
  {\bibfnamefont {X.}~\bibnamefont {Xu}}, \ and\ \bibinfo {author}
  {\bibfnamefont {B.~C.}\ \bibnamefont {Sales}},\ }\href {\doibase
  10.1103/PhysRevMaterials.1.014001} {\bibfield  {journal} {\bibinfo  {journal}
  {Phys. Rev. Materials}\ }\textbf {\bibinfo {volume} {1}},\ \bibinfo {pages}
  {014001} (\bibinfo {year} {2017})}\BibitemShut {NoStop}%
\bibitem [{\citenamefont {Glamazda}\ \emph {et~al.}(2017)\citenamefont
  {Glamazda}, \citenamefont {Lemmens}, \citenamefont {Do}, \citenamefont
  {Kwon},\ and\ \citenamefont {Choi}}]{glamazda2017}%
  \BibitemOpen
  \bibfield  {author} {\bibinfo {author} {\bibfnamefont {A.}~\bibnamefont
  {Glamazda}}, \bibinfo {author} {\bibfnamefont {P.}~\bibnamefont {Lemmens}},
  \bibinfo {author} {\bibfnamefont {S.-H.}\ \bibnamefont {Do}}, \bibinfo
  {author} {\bibfnamefont {Y.~S.}\ \bibnamefont {Kwon}}, \ and\ \bibinfo
  {author} {\bibfnamefont {K.-Y.}\ \bibnamefont {Choi}},\ }\href {\doibase
  10.1103/PhysRevB.95.174429} {\bibfield  {journal} {\bibinfo  {journal} {Phys.
  Rev. B}\ }\textbf {\bibinfo {volume} {95}},\ \bibinfo {pages} {174429}
  (\bibinfo {year} {2017})}\BibitemShut {NoStop}%
\bibitem [{\citenamefont {Chen}\ \emph
  {et~al.}(2021{\natexlab{b}})\citenamefont {Chen}, \citenamefont {Stone},
  \citenamefont {Kolesnikov}, \citenamefont {Winn}, \citenamefont {Shon},
  \citenamefont {Dai},\ and\ \citenamefont {Chung}}]{chen2021crcl}%
  \BibitemOpen
  \bibfield  {author} {\bibinfo {author} {\bibfnamefont {L.}~\bibnamefont
  {Chen}}, \bibinfo {author} {\bibfnamefont {M.~B.}\ \bibnamefont {Stone}},
  \bibinfo {author} {\bibfnamefont {A.~I.}\ \bibnamefont {Kolesnikov}},
  \bibinfo {author} {\bibfnamefont {B.}~\bibnamefont {Winn}}, \bibinfo {author}
  {\bibfnamefont {W.}~\bibnamefont {Shon}}, \bibinfo {author} {\bibfnamefont
  {P.}~\bibnamefont {Dai}}, \ and\ \bibinfo {author} {\bibfnamefont {J.-H.}\
  \bibnamefont {Chung}},\ }\href {\doibase 10.1088/2053-1583/ac2e7a} {\bibfield
   {journal} {\bibinfo  {journal} {2D Materials}\ }\textbf {\bibinfo {volume}
  {9}},\ \bibinfo {pages} {015006} (\bibinfo {year}
  {2021}{\natexlab{b}})}\BibitemShut {NoStop}%
\bibitem [{\citenamefont {Schneeloch}\ \emph {et~al.}(2021)\citenamefont
  {Schneeloch}, \citenamefont {Tao}, \citenamefont {Cheng}, \citenamefont
  {Daemen}, \citenamefont {Xu}, \citenamefont {Zhang},\ and\ \citenamefont
  {Louca}}]{schneeloch2021}%
  \BibitemOpen
  \bibfield  {author} {\bibinfo {author} {\bibfnamefont {J.~A.}\ \bibnamefont
  {Schneeloch}}, \bibinfo {author} {\bibfnamefont {Y.}~\bibnamefont {Tao}},
  \bibinfo {author} {\bibfnamefont {Y.}~\bibnamefont {Cheng}}, \bibinfo
  {author} {\bibfnamefont {L.}~\bibnamefont {Daemen}}, \bibinfo {author}
  {\bibfnamefont {G.}~\bibnamefont {Xu}}, \bibinfo {author} {\bibfnamefont
  {Q.}~\bibnamefont {Zhang}}, \ and\ \bibinfo {author} {\bibfnamefont
  {D.}~\bibnamefont {Louca}},\ }\href@noop {} {\bibfield  {journal} {\bibinfo
  {journal} {arXiv preprint arXiv:2110.10771}\ } (\bibinfo {year}
  {2021})}\BibitemShut {NoStop}%
\bibitem [{\citenamefont {May}\ \emph {et~al.}(2020)\citenamefont {May},
  \citenamefont {Yan},\ and\ \citenamefont {McGuire}}]{may2020}%
  \BibitemOpen
  \bibfield  {author} {\bibinfo {author} {\bibfnamefont {A.~F.}\ \bibnamefont
  {May}}, \bibinfo {author} {\bibfnamefont {J.}~\bibnamefont {Yan}}, \ and\
  \bibinfo {author} {\bibfnamefont {M.~A.}\ \bibnamefont {McGuire}},\ }\href
  {\doibase 10.1063/5.0015971} {\bibfield  {journal} {\bibinfo  {journal}
  {Journal of Applied Physics}\ }\textbf {\bibinfo {volume} {128}},\ \bibinfo
  {pages} {051101} (\bibinfo {year} {2020})}\BibitemShut {NoStop}%
\bibitem [{\citenamefont {Granroth}\ \emph {et~al.}(2010)\citenamefont
  {Granroth}, \citenamefont {Kolesnikov}, \citenamefont {Sherline},
  \citenamefont {Clancy}, \citenamefont {Ross}, \citenamefont {Ruff},
  \citenamefont {Gaulin},\ and\ \citenamefont {Nagler}}]{granroth2010}%
  \BibitemOpen
  \bibfield  {author} {\bibinfo {author} {\bibfnamefont {G.~E.}\ \bibnamefont
  {Granroth}}, \bibinfo {author} {\bibfnamefont {A.~I.}\ \bibnamefont
  {Kolesnikov}}, \bibinfo {author} {\bibfnamefont {T.~E.}\ \bibnamefont
  {Sherline}}, \bibinfo {author} {\bibfnamefont {J.~P.}\ \bibnamefont
  {Clancy}}, \bibinfo {author} {\bibfnamefont {K.~A.}\ \bibnamefont {Ross}},
  \bibinfo {author} {\bibfnamefont {J.~P.~C.}\ \bibnamefont {Ruff}}, \bibinfo
  {author} {\bibfnamefont {B.~D.}\ \bibnamefont {Gaulin}}, \ and\ \bibinfo
  {author} {\bibfnamefont {S.~E.}\ \bibnamefont {Nagler}},\ }\href {\doibase
  10.1088/1742-6596/251/1/012058} {\bibfield  {journal} {\bibinfo  {journal}
  {Journal of Physics: Conference Series}\ }\textbf {\bibinfo {volume} {251}},\
  \bibinfo {pages} {012058} (\bibinfo {year} {2010})}\BibitemShut {NoStop}%
\bibitem [{sup()}]{supple}%
  \BibitemOpen
  \href@noop {} {\enquote {\bibinfo {title} {See supplemental material for
  additional data and analysis.}}\ }\BibitemShut {NoStop}%
\bibitem [{\citenamefont {Toth}\ and\ \citenamefont {Lake}(2015)}]{toth2015}%
  \BibitemOpen
  \bibfield  {author} {\bibinfo {author} {\bibfnamefont {S.}~\bibnamefont
  {Toth}}\ and\ \bibinfo {author} {\bibfnamefont {B.}~\bibnamefont {Lake}},\
  }\href {\doibase 10.1088/0953-8984/27/16/166002/meta} {\bibfield  {journal}
  {\bibinfo  {journal} {Journal of Physics: Condensed Matter}\ }\textbf
  {\bibinfo {volume} {27}},\ \bibinfo {pages} {166002} (\bibinfo {year}
  {2015})}\BibitemShut {NoStop}%
\bibitem [{\citenamefont {Elliot}\ \emph {et~al.}(2021)\citenamefont {Elliot},
  \citenamefont {McClarty}, \citenamefont {Prabhakaran}, \citenamefont
  {Johnson}, \citenamefont {Walker}, \citenamefont {Manuel},\ and\
  \citenamefont {Coldea}}]{elliot2020}%
  \BibitemOpen
  \bibfield  {author} {\bibinfo {author} {\bibfnamefont {M.}~\bibnamefont
  {Elliot}}, \bibinfo {author} {\bibfnamefont {P.~A.}\ \bibnamefont
  {McClarty}}, \bibinfo {author} {\bibfnamefont {D.}~\bibnamefont
  {Prabhakaran}}, \bibinfo {author} {\bibfnamefont {R.}~\bibnamefont
  {Johnson}}, \bibinfo {author} {\bibfnamefont {H.}~\bibnamefont {Walker}},
  \bibinfo {author} {\bibfnamefont {P.}~\bibnamefont {Manuel}}, \ and\ \bibinfo
  {author} {\bibfnamefont {R.}~\bibnamefont {Coldea}},\ }\href {\doibase
  https://doi.org/10.1038/s41467-021-23851-0} {\bibfield  {journal} {\bibinfo
  {journal} {Nature Communications}\ }\textbf {\bibinfo {volume} {12}},\
  \bibinfo {pages} {3936} (\bibinfo {year} {2021})}\BibitemShut {NoStop}%
\bibitem [{\citenamefont {Shivam}\ \emph {et~al.}(2017)\citenamefont {Shivam},
  \citenamefont {Coldea}, \citenamefont {Moessner},\ and\ \citenamefont
  {McClarty}}]{shivam2017}%
  \BibitemOpen
  \bibfield  {author} {\bibinfo {author} {\bibfnamefont {S.}~\bibnamefont
  {Shivam}}, \bibinfo {author} {\bibfnamefont {R.}~\bibnamefont {Coldea}},
  \bibinfo {author} {\bibfnamefont {R.}~\bibnamefont {Moessner}}, \ and\
  \bibinfo {author} {\bibfnamefont {P.}~\bibnamefont {McClarty}},\ }\href
  {https://arxiv.org/abs/1712.08535} {\bibfield  {journal} {\bibinfo  {journal}
  {arXiv preprint arXiv:1712.08535}\ } (\bibinfo {year} {2017})}\BibitemShut
  {NoStop}%
\bibitem [{\citenamefont {Farhi}\ \emph {et~al.}(2014)\citenamefont {Farhi},
  \citenamefont {Debab},\ and\ \citenamefont {Willendrup}}]{reslib}%
  \BibitemOpen
  \bibfield  {author} {\bibinfo {author} {\bibfnamefont {E.}~\bibnamefont
  {Farhi}}, \bibinfo {author} {\bibfnamefont {Y.}~\bibnamefont {Debab}}, \ and\
  \bibinfo {author} {\bibfnamefont {P.}~\bibnamefont {Willendrup}},\ }\href
  {\doibase 10.3233/JNR-130001} {\bibfield  {journal} {\bibinfo  {journal}
  {Journal of Neutron Research}\ }\textbf {\bibinfo {volume} {17}},\ \bibinfo
  {pages} {5} (\bibinfo {year} {2014})}\BibitemShut {NoStop}%
\bibitem [{\citenamefont {Christensen}\ \emph {et~al.}(2008)\citenamefont
  {Christensen}, \citenamefont {Abrahamsen}, \citenamefont {Christensen},
  \citenamefont {Juranyi}, \citenamefont {Andersen}, \citenamefont {Lefmann},
  \citenamefont {Andreasson}, \citenamefont {Bahl},\ and\ \citenamefont
  {Iversen}}]{christensen2008}%
  \BibitemOpen
  \bibfield  {author} {\bibinfo {author} {\bibfnamefont {M.}~\bibnamefont
  {Christensen}}, \bibinfo {author} {\bibfnamefont {A.~B.}\ \bibnamefont
  {Abrahamsen}}, \bibinfo {author} {\bibfnamefont {N.~B.}\ \bibnamefont
  {Christensen}}, \bibinfo {author} {\bibfnamefont {F.}~\bibnamefont
  {Juranyi}}, \bibinfo {author} {\bibfnamefont {N.~H.}\ \bibnamefont
  {Andersen}}, \bibinfo {author} {\bibfnamefont {K.}~\bibnamefont {Lefmann}},
  \bibinfo {author} {\bibfnamefont {J.}~\bibnamefont {Andreasson}}, \bibinfo
  {author} {\bibfnamefont {C.~R.}\ \bibnamefont {Bahl}}, \ and\ \bibinfo
  {author} {\bibfnamefont {B.~B.}\ \bibnamefont {Iversen}},\ }\href {\doibase
  /10.1038/nmat2273} {\bibfield  {journal} {\bibinfo  {journal} {Nature
  materials}\ }\textbf {\bibinfo {volume} {7}},\ \bibinfo {pages} {811}
  (\bibinfo {year} {2008})}\BibitemShut {NoStop}%
\bibitem [{\citenamefont {Xiao}\ \emph {et~al.}(2022)\citenamefont {Xiao},
  \citenamefont {Ma}, \citenamefont {Bryan}, \citenamefont {Fu}, \citenamefont
  {Mann}, \citenamefont {Winn}, \citenamefont {Abernathy}, \citenamefont
  {Hermann}, \citenamefont {Khanolkar}, \citenamefont {Dennett} \emph
  {et~al.}}]{xiao2022}%
  \BibitemOpen
  \bibfield  {author} {\bibinfo {author} {\bibfnamefont {E.}~\bibnamefont
  {Xiao}}, \bibinfo {author} {\bibfnamefont {H.}~\bibnamefont {Ma}}, \bibinfo
  {author} {\bibfnamefont {M.~S.}\ \bibnamefont {Bryan}}, \bibinfo {author}
  {\bibfnamefont {L.}~\bibnamefont {Fu}}, \bibinfo {author} {\bibfnamefont
  {J.~M.}\ \bibnamefont {Mann}}, \bibinfo {author} {\bibfnamefont
  {B.}~\bibnamefont {Winn}}, \bibinfo {author} {\bibfnamefont {D.~L.}\
  \bibnamefont {Abernathy}}, \bibinfo {author} {\bibfnamefont {R.~P.}\
  \bibnamefont {Hermann}}, \bibinfo {author} {\bibfnamefont {A.~R.}\
  \bibnamefont {Khanolkar}}, \bibinfo {author} {\bibfnamefont {C.~A.}\
  \bibnamefont {Dennett}},  \emph {et~al.},\ }\href@noop {} {\bibfield
  {journal} {\bibinfo  {journal} {arXiv preprint arXiv:2202.11041}\ } (\bibinfo
  {year} {2022})}\BibitemShut {NoStop}%
\end{thebibliography}
%\bibliographystyle{apsrev4-1}

\providecommand{\noopsort}[1]{}\providecommand{\singleletter}[1]{#1}%

\clearpage

\addtolength{\oddsidemargin}{-0.75in}
\addtolength{\evensidemargin}{-0.75in}
\addtolength{\topmargin}{-0.725in}

\newcommand{\addpage}[1] {
\begin{figure*}
  \includegraphics[width=8.5in,page=#1]{Supplemental.pdf}
\end{figure*}
}

\addpage{1}
\addpage{2}
\addpage{3}
\addpage{4}
\addpage{5}
\addpage{6}
\addpage{7}

\end{document}